\documentclass[a4paper,12pt]{article}
\usepackage{amsmath,amssymb}
\usepackage[dvipdfmx]{graphicx}

\newcommand{\be}{\begin{equation}}
\newcommand{\ee}{\end{equation}}
\newcommand{\f}{\frac}
\newcommand{\s}{\sqrt}
\newcommand{\p}{\partial}

\newcommand{\om}{\mathcal{O}_{n}}
\newcommand{\bea}{\begin{eqnarray}}
\newcommand{\eea}{\end{eqnarray}}
\newcommand{\ba}{\begin{align}}
\newcommand{\ea}{\end{align}}

\begin{document}

\begin{titlepage}
\begin{flushright}
IPMU13-0221

YITP-13-116
\end{flushright}

\vspace{.4cm}
\begin{center}
\noindent{\Large \textbf{Two dimensional quantum quenches and holography}}\\
\vspace{2cm}
Tomonori Ugajin $^{a,b,}$\footnote{e-mail:tomonori.ugajin@ipmu.jp}

\vspace{1cm}
  {\it
 $^{a}$Kavli Institute for the Physics and Mathematics of the Universe,\\
University of Tokyo, Kashiwa, Chiba 277-8582, Japan\\
$^{b}$Yukawa Institute for Theoretical Physics,
Kyoto University, \\
Kitashirakawa Oiwakecho, Sakyo-ku, Kyoto 606-8502, Japan,\\
\vspace{0.2cm}
 }
\vskip 2em
\end{center}

\vspace{.5cm}

\begin{abstract}
We propose a  holographic realization of quantum quenches in two dimensional 
conformal field theories. In particular, we discuss time evolutions of  holographic entanglement entropy in these backgrounds and compare them with CFT results. The key ingredient of the construction is an introduction 
of a spacetime boundary  into bulk geometries, which is the gravity counterpart of a boundary state in the dual CFT. We consider several examples, including local quenches and an inhomogeneous quench which is  dual  to fusion of two black string into the third one.  
 
\end{abstract}
\end{titlepage}

\tableofcontents

\section{Introduction}

Entanglement entropy is a useful quantity to characterize not only vacuum, but also excited states.
It has been shown to be an efficient measure to probe thermalization processes since it can be 
regarded as a non-equilibrium generalization of thermal entropy. 

%
%
%
%
%
%

An interesting toy model of thermalization in two dimensions is proposed in \cite{cc0503}, known as global quenches. In a global quench, one prepares a ground state of a gapped Hamiltonian, then excites the system by suddenly changing parameters in the Hamiltonian so that the system becomes gappless, while keeping the translational invariance. The resulting excited state is supposed to have both translational and conformal invariance below the energy scale of the quench, therefore it is well approximated by a boundary state of the gapless theory ($\rm{CFT}_{2}$). The observation about the initial state makes it possible to compute the time evolution of entanglement entropy under the quench by calculating  correlation functions of  a twisted operator on a strip. The resulting entanglement entropy grows linearly in time and subsequently it is thermalized. The result is interpreted via entangling pairs of quasi particles propagating in opposite direction.  

Holography \cite{Maldacena:1997re,Witten:1998qj,Gubser:1998bc} is also a useful tool to study time dependent processes of  dual field theories. In particular, gravity descriptions of global quenches have been discussed. See for example \cite{AbajoArrastia:2010yt,Aparicio:2011zy,Albash:2010mv,
Balasubramanian:2010ce,Balasubramanian:2011ur,Liu:2013iza,Liu:2013qca}.
In the works,
The behavior of entanglement entropy in these quenches was reproduced by 
applying the holographic formula \cite{Ryu:2006bv,Hubeny:2007xt} to Vidya spacetimes which are toy models of  gravitational collapses. We also refer \cite{Buchel:2013lla,Buchel:2013gba} for holographic studies of an another type
of global quench induced by sudden deformation of  theories by relevant operators .

Remarkably, in the recent paper \cite{Hartman:2013qma} it is shown that the linear growth of entanglement entropy in these quenches is actually intrinsic to black branes which are final states of the gravitational collapses,
and related to the growth  of the interior region of these black branes in a particular time slice. As a generalization, in \cite{Shenker:2013pqa}  the time evolutions of mutual informations 
are analyzed in the model where an energy flux is ingoing to the horizon an eternal black hole.
 

Actually there are variety of quantum quenches in two dimensional  CFTs.   
For example one can consider quenches where one injects the energy only at one 
point of the initial time slice. These quenches are called local quenches \cite{cc0708}. The time evolution of entanglement entropy was computed by calculating certain correlation functions on
a plane with slits. An attempt for a holographic realization has been made in \cite{Nozaki:2013wia}.

In general, there is one to one correspondence between quantum quenches in 2d CFTs and Riemann surfaces with
boundaries. Certain correlation functions on these Riemann surfaces are related to time evolutions of entanglement entropy under corresponding quenches. The dictionary relates the energy density $\epsilon(x)$ injected at each point $x$ of the initial time slice in the quench to the locations of the boundaries $x\pm  \f{i}{\epsilon (x)} $ in the Riemann surface.

In this paper we would like to explore a description of  holographic duals of general quantum quenches in 2d $CFT$s. 
The main idea is the introduction of a spacetime boundary in the bulk.This is motived by the holographic realization of CFT on a manifold with boundary (BCFT) \cite{Karch:2000ct,DeWolfe:2001pq,Takayanagi:2011zk, Fujita:2011fp, Nozaki:2012qd}.   We also refer for a recent  discussion of entanglement entropy in the presence of boundaries \cite{Jensen:2013lxa}.
When a spacetime boundary is present in the bulk, there is an extremal surface in the 
system which is attaching to the spacetime boundary in addition to usual one which connects the boundary of the subsystem.   We call the extremal surface attaching on the spacetime boundary as disconnected surface and see that introduction of it is essential to reproduce early time behaviors of
entanglement entropy under quantum quenches holographically.

The general prescription make a systematic construction of gravity duals of 2d quantum quenches possible, including the global quench/BTZ black string correspondence \cite{Hartman:2013qma} as 
a specific example. Moreover we discuss that the gravity dual of  a certain inhomogeneous quench involves fusion of two black strings into the third black string.  The time evolution of  entanglement entropy in the quench is explained in terms of dynamics of these three black strings. We also comment how the evolutions of entanglement entropy in local quenches are reproduced from  AdS shock wave metrics. Furthermore, our prescription  resolves a certain mismatch between holographic and CFT result found in \cite{Nozaki:2013wia}
by virtue of the spacetime boundary we introduced.

The organization of the paper is as follows. In section \ref{sec:rev} we review several quantum quenches in 2d CFTs with 
emphasis on their relation to Riemann surfaces with boundaries. In section \ref{sec:holq} we explain a general prescription to holographically realize 
2d quantum quenches by making use of a spacetime boundary. In section \ref{sec:examples} we deal with two examples, namely gravity dual of local quench and inhomogeneous quench. We interpret behaviors of  entanglement entropy under these quenches  from corresponding bulk geometries.

\section{Review of various quenches and time evolutions of entanglement entropy}\label{sec:rev}
In this section we review various quantum quenches in two dimensional CFTs and time evolutions 
of  entanglement entropy during these processes.

In a typical quantum quench, we prepare a ground state of a gapped Hamiltonian at the initial time. We then make the system gapless suddenly. In the process, the prepared ground state of the gapped Hamiltonian becomes a excited state of the
gappless Hamiltonian, thus it evolves non trivially. We want to compute time dependent  entanglement entropy of the excited state.  

%
Entanglement entropy $S_{A}(t)$ of a generic state $ |\psi \rangle$ for a subsystem $A$ can be computed by the replica method.
\be
S_{A}(t)=-\f{\p}{\p n}\Big|_{n=1} {\rm tr} \rho_{A}^{n}(t).
\ee

The reduce density matrix $\rho_{A}$ for a subsystem A is defined as
\be
\rho_{A}= {\rm tr}_{A^{c}}[ \rho] \qquad \rho= e^{i Ht}|\psi \rangle \langle \psi |e^{-i Ht},
\ee
where $A^{c}$ denotes the complement of a subsystem $A$. 
Actual computations are performed in the Euclidean signature $t \rightarrow -i \tau $.

 In two dimensional CFTs on a Riemann surface $\Sigma$, computations of traces of  reduced density matrices boil down to computation of  correlation functions of a twisted operator on $\Sigma$ \cite{Calabrese:2004eu}. For example, 
when the subsystem A is an interval $[l_{1},l_{2}]$,the relation is
\be
 {\rm tr} \rho_{A}^{n}(\tau) =\langle \mathcal{O}(l_{1},\tau) \mathcal{O}(l_{2},\tau) \rangle_{\Sigma}.
\ee 
The conformal weights of the operator are $\Delta_{n}=\bar{\Delta}_{n}=\f{c}{24}\left(n-\f{1}{n}\right)$, where $c$ is the central charge of the CFT. Generalizations to multi intervals are straightforward.
 As we will see below, by changing the background Riemann surface $\Sigma$ one can change the way to quench the system. Since All of these relevant Riemann surfaces can be mapped to a half plane via conformal maps, we can determine
correlation functions of the twisted operator on these Riemann surfaces. 
%
We then analytically continue the imaginary time $\tau \rightarrow it$ and derive real time evolution of entanglement entropy. 

\subsection{Global quenches}

In this subsection we review global quenches. In these quenches we  excite each points of the spatial direction  equally.
The resulting initial states $|\psi \rangle$ are translational and conformal invariant below  the energy scale of the quench $4/\beta$, thus they can be written by  boundary states $|B \rangle$ 
of the CFT as $|\psi \rangle = e^{-\f{\beta}{4} H}|B \rangle$. A trace of a reduced density matrix of these excited states can be computed by  a partition function of the CFT on a strip whose width is $\f{\beta}{2}$ with corresponding boundary conditions on boundaries and branch cut along the subsystem A . The partition function is equivalent
to a correlation function of a twisted operator, which can be computed by mapping the strip to a half plane.
The strip and a half plane are related via  
\be
w= \f{\beta}{2\pi}\log z. \label{eq:cmp}
\ee
%
The two point function of the twisted operator on a half plane have following form
\be
\langle \om (w_{1}) \om (w_{2} )\rangle_{HP}=\left(\f{|w_{1}+\bar{w_{2}}||w_{2}+\bar{w_{1}}|}{|w_{1}-w_{2}||\bar{w_{1}}-\bar{w_{2}}||w_{1}+\bar{w_{1}}||w_{2}+\bar{w_{2}}|}\right)^{2\Delta_{n}}. \label{eq:2point}
\ee
Generically the two point function contains a function that depends on cross ratio. However in sufficiently early time $t \ll |l_{2}-l_{1}|$ or late time $t \gg |l_{2}-l_{1}|$, the two point function is factorized. Therefore contributions from the function can be negligible in these limit. See for example \cite{cc0711}  
and explicit calculations \cite{Takayanagi:2010wp}. Below we assume this.
The $  {\rm tr} \rho_{A}^{n}(t)$ on the strip when the subsystem A is a interval is given by applying the conformal map (\ref{eq:cmp}),
\be
 {\rm tr} \rho_{A}^{n}(t)= \langle \om (\f{l}{2}+it) \om (-\f{l}{2}+it )\rangle_{{\rm strip}}=c_{n}\left(\f{ \sinh (\f{2\pi l}{\beta})+\cosh(\f{4\pi t}{\beta})}{\sinh ^2({\f{\pi l}{\beta}})\cosh ^2(\f{2\pi t}{\beta})} \right)^{2\Delta_{n}},
\ee
where $c_{n}$ is a constant which does not depend on $l,t$ or $\beta$.
Because of homogeneity of the quench, one can take the subsystem A to be $[-\f{l}{2},\f{l}{2}]$ without loss of generality.
If we take the high temperature limit $\beta \rightarrow 0$, entanglement entropy behaves like
\be
S_{A}(t)= \frac{2\pi ct}{3\beta}\theta ( \f{l}{2}-t) +\f{\pi c l}{3\beta}\theta (t- \f{l}{2})
\ee
Where the $\theta (x)$ is the step function.  From the expression  we see that when $t \leq \f{l}{2}$
$S_{A}(t)$ grows linearly in time , and when $t \geq \f{l}{2}$ it takes thermal value with the inverse temperature $\beta$.
The stress energy tensors on the strip is given by 
\be
\langle T_{zz} \rangle=\f{c}{24}\left(\frac{2\pi}{\beta}\right)^2 \qquad \langle \bar{T}_{\bar{z}\bar{z}} \rangle=\f{c}{24}\left(\frac{2\pi}{\beta}\right)^2 \label{eq:stg}
\ee
Thus the width of the strip $\beta$ can be identified as inverse of effective temperature induced by the quench.


\subsection{Inhomogeneous quench} 

Having identified the width of the strip to inverse of effective temperature we add,
one can also consider various generalizations of global quenches where the energy density $1/\beta $ one adds depend on the position in the spacial direction $x$. These quenches are called inhomogeneous quenches \cite{sc0808}. To treat them we need to deal with
wavy strips whose width depend on the position in the spacial direction. 

Let us consider a wavy strip extending 
along the $x$ direction, and at each $x$, width is given by $\f{1}{2}(\beta+h(x))$.
If inhomogeneity is infinitesimally small, $ h(x) \ll \beta$, the conformal map which maps the wavy strip with coordinate $z$ to the ordinary strip with width
$\f{\beta}{2}$ and coordinate $w$ is given by $w=z-f(z) \label{eq:back}$, where the $f(z)$ is
\be
f(z)=\int^{\infty}_{-\infty}F(s-z)h(s)ds.
\ee
The karnel $F(s-z)$ is 
\be
F(z)=\f{1}{2\beta}\left(\tanh \f{2\pi z}{\beta}+1 \right),
\ee
the detail of the construction is in \cite{sc0808}.
One can get a simpler expression when  $z$ lies in the real axis of the wavy strip and $\beta \rightarrow 0$ limit is taken, because the kernel reduces to a step function. Then the $f(x)$ reduces to
\be
f(x)=\f{1}{\beta} \int ^{x}_{-\infty} h(s)ds \label{eq:backrelation}.
\ee
One can  compute correlation functions of the twisted operator in the 
wavy strip by using the map $w=z-f(z)$ with (\ref{eq:backrelation}) for arbitrary $h(x)$ .The final expression of  the entanglement entropy in high temperature 
limit is 


\begin{align}
S(l_{1},l_{2},t) &=S_{0}(l_{1},l_{2},t) \label{eq:inhee} \\ \nonumber 
&-\f{c\pi}{6\beta^2}\theta(\frac{l_{2}-l_{1}}{2}-t)\left[\int^{l_{1}+t}_{l_{1}-t}+\int^{l_{2}+t}_{l_{2}-t} h(s) ds \right]\\ \nonumber 
&-\f{c\pi}{6 \beta^2}\theta(t-\frac{l_{2}-l_{1}}{2})\left[ \int^{l_{2}-t}_{l_{1}-t}+\int^{l_{2}+t}_{l_{1}+t} h(s) ds \right] \\ \nonumber 
&+\f{c}{12} \left(h(l_{1}-t)+h(l_{1}+t)+h(l_{2}-t)+h(l_{2}+t) \right)
\end{align}

Where the $S_{0}(x_{1},x_{2},t) $ denote the time evolution of entanglement entropy in the global quench.
The second  and the third term can be interpreted by the quasi particle picture. 
The quasi particles which contribute to entanglement at fixed time $t< \f{l_{1}-l_{2}}{2}$ are  lie in the segment $[ l_{1}-t, l_{1}+t]$ and $[ l_{2}-t, l_{2}+t]$ at the initial time. 
The total contributions of them to entanglement entropy is 
\be
\f{c\pi}{6}\int^{l_{1}+t}_{l_{1}-t}+\int^{l_{2}+t}_{l_{2}-t} \f{ds}{\beta +h(x)}. 
\ee
If we set $h(x)=0$, the expression reproduces the linear grow of the global quench in the early time. Thus one can regard it  a direct generalization of the global quench. Since we are assuming $h(x) \ll \epsilon$ we get the second term of (\ref{eq:inhee} ).
Third term can be interpreted by a similar manner.

The  time evolutions of stress energy tensors are

\begin{align}
\langle T_{++}(x,t) \rangle&=(1-2f'(x+t)) \f{c}{24}\left(\frac{2\pi}{\beta}\right)^2 +\f{c}{24}f'''(x+t) \label{eq:inhem} \\ \nonumber 
\langle T_{--}(x,t) \rangle& =(1-2f'(x-t)) \f{c}{24}\left(\frac{2\pi}{\beta}\right)^2 +\f{c}{24}f'''(x-t) . \\ \nonumber
\end{align}

\subsection{Local  quenches}
We can consider a quench where we inject energy density $1/\epsilon$ only at one point of the initial time slice\cite{cc0708}. In the quench 
the corresponding Riemann surface is the plane with two slits which extend $[-i \infty,-i\epsilon]$ and $[i\epsilon,i\infty]$. The Riemann surface is 
mapped to the half plane  via \cite{cc0708}
\be
w =\f{z}{\epsilon}+\sqrt{\left(\f{z}{\epsilon}\right)^2+1} \label{eq:maplq}
\ee

\begin{figure}
\begin{minipage}{0.5\hsize}
\begin{center}
   \includegraphics[width=50mm]{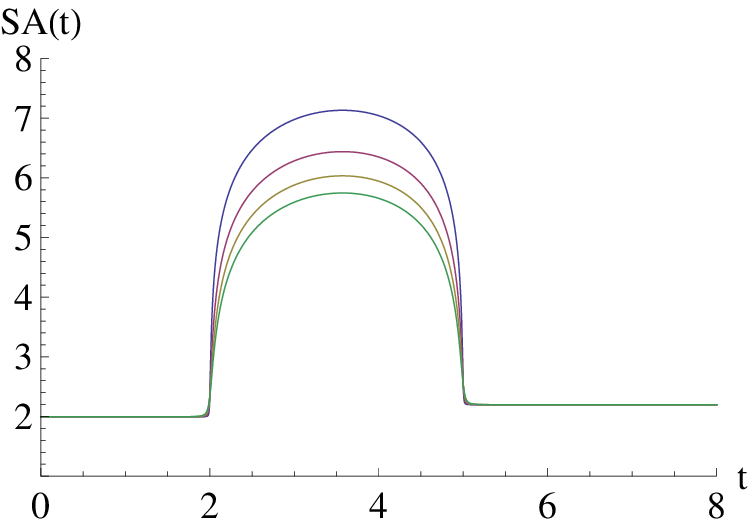}
\end{center}
\end{minipage}
\begin{minipage}{0.5\hsize}
\begin{center}
\includegraphics[width=50mm]{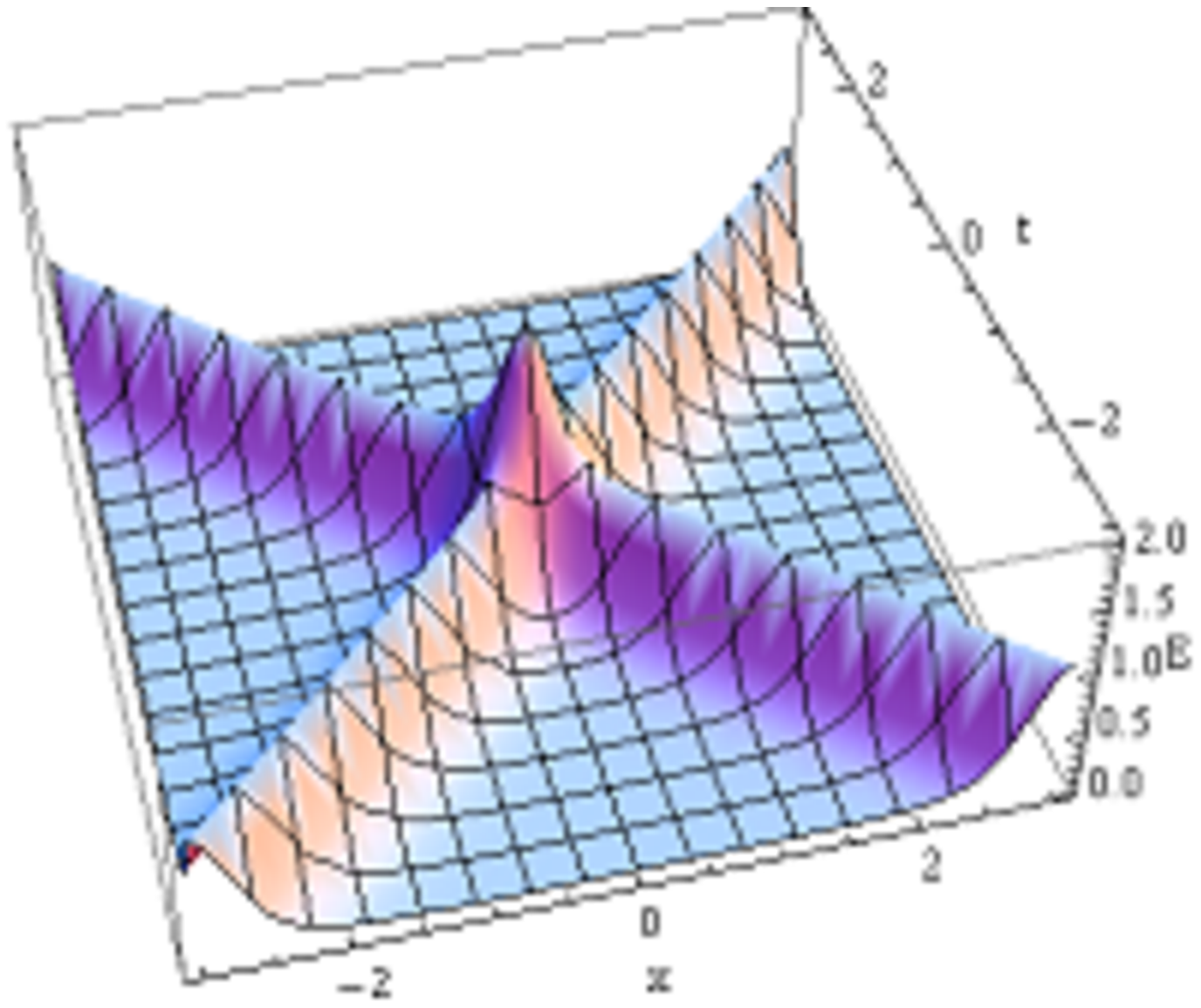}
\end{center}
\end{minipage}
\caption{Left:Plot of the time  evolution of entanglement entropy under local quenches for various values of $\epsilon$.As we decrease $\epsilon$, maximum value of entanglement entropy become larger. We take $l_{1}=2,l_{2}=5, c=6$. 
Right: Plot of the time evolution of  energy density $\langle T_{tt}(x,t) \rangle$ during the quench. We take $\epsilon=\f{1}{2}, c=2$.}
\label{fig:lqeecft}
\end{figure}

Suppose we take the subsystem  A  to be a finite interval $[ l_{1},l_{2}]$. The time evolution of the 
entanglement entropy under the quench can be computed by substituting the analytic continuation of the map (\ref{eq:maplq}) into the general formula (\ref{eq:cfteegen}) of the Appendix.
\begin{align}
S_{A}(l_{1},l_{2},t)&= \f{c}{6}\log \left[ \left(2l_{1}+\s{(l_{1}+t)^2+\epsilon^2}+\s{(l_{1}-t)^2+\epsilon^2}\right)\left(l_{1} \leftrightarrow l_{2}\right)\right] \label{eq:lqCFT} \\  \nonumber
&+ \f{c}{6}\log \left[ \left((l_{2}-l_{1})+\s{(l_{2}+t)^2+\epsilon^2}-\s{(l_{1}+t)^2+\epsilon^2}\right)\left(t \leftrightarrow -t \right)\right]\\ \nonumber
&- \f{c}{6}\log \left[ \left((l_{1}+l_{2})+\s{(l_{1}+t)^2+\epsilon^2}+\s{(l_{2}-t)^2+\epsilon^2}\right)\left(t \leftrightarrow -t \right)\right]\\ \nonumber
&-\f{c}{12} \log \left[a_{LQ}^2\left( 1+\f{l_{1}+t}{\s{(l_{1}+t)^2+\epsilon^2}}\right) \left( 1+\f{l_{1}-t}{\s{(l_{1}-t)^2+\epsilon^2}}\right)\left(l_{1} \leftrightarrow l_{2}\right)\right]\\ \nonumber
\end{align}
Figure \ref{fig:lqeecft} show the plot of the time evolution of entanglement entropy  for several values of $\epsilon$. 
At the late time $t \rightarrow \infty$ the expression settles down to the vacuum value
\be
S_{A}(l_{1},l_{2},t) \rightarrow  \f{c}{3} \log \f{|l_{1}-l_{2}|}{a_{LQ}}, \qquad  t \rightarrow \infty .\label{eq:lqcft1}
\ee

The early time behavior of (\ref {eq:lqCFT}) depends on the sign of $l_{1}$. In the high energy limit $\epsilon \rightarrow 0$, 
When $0<l_{1}<l_{2}$,
\be
S_{A}(t)= \f{c}{6} \log \left[ \f{8l_{1}l_{2}(l_{1}-l_{2})^2}{(l_{1}+l_{2})^2}\right],  \quad t \sim 0 \label{eq:lqcft2}
\ee
When 
$l_{1}<0<l_{2}$,
\be
S_{A}(t)=\f{c}{6} \log \f{4|l_{1}|l_{2}}{a_{LQ}^2}, \quad t \sim 0 .\label{eq:lqcft3}
\ee

Time dependence of entanglement entropy $S_{A}(t)$ in the process can be explained by the quasi particle picture. Let us consider the $l_{1}<0<l_{2}$ case for example.
In the local quench, an entangled pair of quasi particles with speed of light is created at $t=0, x=0$ where the quench happens.
When $0< t< \min \{|l_{1}|,l_{2} \}$, the entangled pair completely lie inside the subsystem A, thus entanglement between the subsystem A and its complement  does not increase.
When $\min \{|l_{1}|,l_{2}\}  \leq t <\max \{ |l_{1}| ,l_{2} \}$ one of the quasi particle lie outside of the subsystem A. Because of this, entanglement between A and its complement 
increases, and $S_{A}$ shows nontrivial time dependence. At $\max \{|l_{1}| ,l_{2} \} \leq t $, both quasi particles  lie outside of the subsystem A, $S_{A}$
does not increase.


\section{A holographic realization of quantum quenches} \label{sec:holq}
In this section, we would like to discuss a way to 
construct a gravity dual of  quantum quenches in 2d CFTs.
 In the 2d CFT side, according to the dictionary we saw in the previous section, a quantum quench  
 is specified by an associated Riemann surface  $\Sigma$ with boundaries. Correlation functions on $\Sigma$, therefore entanglement entropy  can be computed by BCFT techniques. The Riemann surface is mapped to the half plane by an appropriate conformal map $f(z)$. By analytically continuing the set up one obtains time evolution. 

Below we explain gravity descriptions of these CFT set up. This is achieved by introducing a spacetime boundary in the bulk. We also discuss extremal surfaces in the dual gravity system.  We emphasize that in the system there is an additional extremal surface which end on the spacetime boundary we introduced (disconnected surface).
\subsection{A gravity description of CFT on half plane}
It is known that CFT on a half plane have a dual gravity description \cite{Karch:2000ct,DeWolfe:2001pq,Takayanagi:2011zk,Fujita:2011fp,Nozaki:2012qd}.
 The concrete system is a Poincare $AdS_{3}$ with a codimension one brane 
which is located  at an appropriate $AdS_{2}$ submanifold whose boundary coincide with that of half plane. See figure \ref{fig:probebrane}. The configuration keeps $SL(2,R)$ symmetry of the original $SL(2,R) \times SL(2,R)$ of the poincare $AdS_{3}$. Note that similar symmetry breaking pattern appears in CFTs on the half plane. The prescription passes some 
nontrivial checks, for example it reproduces correct form of one point function of a operator in the presence of the boundary in CFT side \cite{DeWolfe:2001pq,Fujita:2011fp}.

To begin, it is convenient to use $AdS_{2}$ foliation of $AdS_{3}$ 
\be
ds^2= d \rho^2 +\cosh ^2 \rho \left( \f{-dT^2+dy^2}{y^2} \right).
\ee
Via a coordinate transformation
\be
Z=\f{y}{\cosh \rho} \qquad X=y\tanh \rho,
\ee
we obtain a Poincare metric,
\be
ds^2=\f{-dT^2+dX^2+dZ^2}{Z^2} \label{eq:poincare}.
\ee
Since the brane we introduced is a codimension one object, its back reaction 
can be treated via following junction condition\cite{Takayanagi:2011zk,Fujita:2011fp}
\be
K_{\mu \nu}-K h_{\mu \nu}=\mathcal{T}h_{\mu \nu}, \label{eq:junction}
\ee
with a boundary condition that demand the brane end at the boundary of the half plane on the boundary of the $AdS_{3}$. The equation determines the location
of the brane.
In the equation, $ h_{\mu \nu}$ denotes the induced metric on the spacetime boundary, $K_{\mu \nu}$ is the extrinsic curvature of it, and $K=K^{\mu}_{\mu}$. 
$\mathcal{T}$ is the tension of the brane.  In the back reacted description, this 
the brane can be regarded as a spacetime boundary. The bulk region enclosed
by the space time boundary and $AdS_{3}$ boundary is proposed to be  holographic duals of BCFTs.

The solution of (\ref{eq:junction}) which satisfy the boundary condition is given by $\rho=\rho_{*}$(constant) plane with
\be
\mathcal{T}=\tanh \rho_{*}.
 \ee
By changing $\mathcal{T}$, one can change boundary state of dual BCFT. In the corresponding  entanglement entropy, this only change a constant term that  does not depend on subsystem we take.\footnote{This quantity is called as boundary entropy \cite{al}. See \cite{Azeyanagi:2007qj} for a holographic calculation.}   Therefore  
Below we set $\mathcal{T}=0$, then the spacetime boundary in the Poincare $AdS_{3}$ is located at $X=0$.

%
  \subsection{The bulk extension of Lorentzian conformal maps} 
The bulk extension of a boundary Lorentzian conformal map  $f_{\pm}(x^{\pm})$ has also been known \cite{Roberts:2012aq}. 
The explicit map is 
\be
W_{\pm}=f_{\pm}(x^{\pm})+\f{2z^2 f_{\pm}'^2 f_{\mp}''}{8f'_{\pm}f'_{\mp}-z^2f''_{\pm}f''_{\mp}} \qquad Z=z \f{(4f'_{+}f'_{-})^{\f{3}{2}}}{8f'_{+}f'_{-}-z^2f''_{+}f''_{-}}, \label{eq:blext}
\ee
where $W^{\pm}=X \pm T$.
The map transforms the 
Poincare metric (\ref{eq:poincare})
to the metric,
\be
ds^2= L_{+}dx_{+}^2+L_{-}dx_{-}^2 +\left(\f{2}{z^2}+\f{z^2}{2}L_{+}L_{-}\right)dx_{+}dx_{-} +\f{dz^2}{z^2}. \label{eq:met}
\ee

where the $L_{\pm}$ are Schwarzian derivatives of the boundary conformal map 

\be
L_{\pm}=\f{3f''^2_{\pm}-2f'_{\pm} f_{\pm}'''}{4f'^2_{\pm}}
\ee
The holographic stress tensor \cite{Balasubramanian:1999re} of the metric is given by 
\be
T_{\pm \pm}=\f{1}{8\pi G_{N}} L_{\pm} \qquad T_{\pm \mp}=0,
\ee 
which are expected from the boundary point of view. 

Combining these two ingredients, a gravity dual of a quantum quench is given by pulling back the Lorenzian $AdS_{3}$ with a spacetime boundary at $X=0$
 by the bulk extension of the Lorentzian conformal map associated with the quench.

\subsection{Extremal surfaces in holographic systems}
 The time evolution of entanglement entropy for a subsystem A in a quench is holographically derived by finding extremal surfaces
in the dual gravity configuration which are anchored to the boundaries of  of the subsystem A.  In the Poincare $AdS_{3}$ with the spacetime boundary at $X=0$, one find  two extremal surfaces which satisfy the boundary condition. Suppose the boundary of the subsystem A is given by $P_{1}(X_{1},T_{1})$ and $P_{2}(X_{2},T_{2})$ 
\begin{figure}
\begin{center}
\includegraphics[width=50mm]{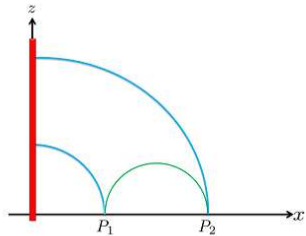}
\end{center}
\caption{Sketch of the spacetime boundary in the Poincare $AdS_{3}$ (thick red line)  at $X=0$ and extremal surfaces in the system.}
\label{fig:probebrane}
\end{figure}
One extremal surface is connecting $P_{1}$ and $P_{2}$, the length of which is given by 
\be
 L_{c}=\log \frac{\sqrt{(X_{1}-X_{2})^2-(T_{1}-T_{2})^2}}{a_{AdS_{3}}} \label{eq:NBAS},
\ee
where $a_{ AdS_{3}}$ denote the UV cut off. Below we call it as connected surface.
The  other surface consist of 2 pieces of disconnected surfaces, one of which is connecting
$P_{1}$ and the spacetime boundary at $X=0$ the other of which is connecting $P_{2}$ and the boundary. The total  length of 
the geodesics is  
\be
 L_{dc}=\f{1}{2}\log \f{X_{1}}{a_{ AdS_{3}}}+ \f{1}{2}\log\f{ X_{2}}{a_{ AdS_{3}}}, \label{eq:BAS}
\ee
below we call it as disconnected surface. See figure \ref{fig:probebrane}.
Extremal surfaces in a gravity dual of a quantum quench are obtained by pulling back extremal surfaces in the Poincare $AdS_{3}$ via (\ref{eq:blext}) .
Holographic entanglement entropy is given in terms of actual minimal surface among them  \cite{Ryu:2006bv,Hubeny:2007xt,Headrick:2010zt},
\be
S_{HEE}=\f{1}{4G_{N}} \min \{L_{c},L_{dc}\},
\ee
where $G_{N}$ is the Newton constant in three dimensions.
As we will see below, there are phase
transitions between these two surfaces in general.

%
%
%
%
%

\subsection{Entanglement Entropy for infinite interval}

When the subsystem A is an infinite interval $[ -\infty,l]$, Time evolutions of  entanglement entropy in the CFT  side
precisely agree with holographic entanglement entropy computed from the disconnected surface for arbitrary quantum quench.
Entanglement entropy of the CFT side is
\be
S_{\rm{CFT}}=\f{c}{6}\log \f{W^{+}(l+t)+W^{-}(l-t)}{a_{Q}^2 \s{\f{dW^{+}}{dx^{+}}(l+t) \f{dW^{-}}{dx^{-}}(l-t)}}.
\ee
Holographic entanglement entropy computed form the disconnected  surface is  
\be
S_{\rm{HEE}}=\frac{c}{6} \log \f{W^{+}(l+t)+W^{-}(l-t)}{a_{ AdS_{3}}}, \quad c=\f{3}{2G_{N}}.
\ee
Since $a_{ AdS_{3}}$ and the location of  the UV cut off of the gravity dual of the quantum quench $a_{Q}$ is related via 
\be
a_{ AdS_{3}}^2=a_{Q}^2\s{\f{dW^{+}}{dx^{+}}(l+t) \f{dW^{-}}{dx^{-}}(l-t) },
\ee 
therefore they precisely  agree. Note that the introduction of the spacetime boundary  and the disconnected surface attaching on it is essential 
to get the correct behavior of  entanglement entropy  holographically.
Below we  consider holographic entanglement entropy for finite intervals.
\section{Application of the construction to various quenches} \label{sec:examples}
\subsection{ global quenches}

As a check, we apply the construction to global quenches. The gravity dual was discussed
in\cite{Hartman:2013qma}. Since one injects energy homogeneously along the spatial direction in these quenches,  corresponding Riemann surfaces are strips. The analytic continuation of the associated conformal  map are

\be
W^{+}= e^{\f{2\pi}{\beta}(x+t)} \qquad W^{-}=e^{\f{2\pi}{\beta}(x-t)} \label{eq:bdcon}
\ee
 where $\beta$ denotes twice of the width of the strip.
One find the gravity dual of the quench via the prescription mentioned in the previous section.
The resulting geometry is nothing but a ordinary BTZ black string with inverse temperature $\beta$.


Next we summarize the time evolution of holographic entanglement entropy.
The contribution of the disconnected surface to  holographic entanglement entropy is 
\be
S_{dc}= \f{c}{3}\log \f{\beta}{\pi a_{Q}}\cosh \frac{2\pi}{\beta}t.
\ee

The contribution of the connected surface is
\be
S{c}= \f{c}{3}\log  \f{\beta}{\pi a_{Q}}\sinh \frac{\pi}{\beta}l,
\ee
 hence there is a phase transition when $t=\f{L}{2}$ in the high temperature limit $\beta \rightarrow 0$.

Our set up of global quenches is closely related 
to that of \cite{Hartman:2013qma}. In the work, they consider BTZ black strings with
 an end of the world brane which is located inside of the horizon $R=\f{U+V}{2}=0$, where $U,V$ are Kruskal coordinates of  these black strings .
By using the map between a BTZ black string and a Poincare $AdS_{3}$,
we can check that the spacetime boundary in the Poincare $AdS_{3}$ is mapped to the end of the world brane they consider in the BTZ black string. Therefore our 
prescription naturally reproduces the result in \cite{Hartman:2013qma}.
\subsection{Infinitesimally inhomogeneous quenches}
Here we would like to discuss a holographic realization of  infinitesimally inhomogeneous quenches \cite{sc0808}  which we reviewed in section 2. 
The metric of the gravity dual is derived by substituting the $\langle T_{\pm \pm} \rangle$ of the inhomogeneous quench (\ref{eq:inhem}) into the 
general metric (\ref{eq:met}). The spacetime can be regarded as BTZ black string with a small perturbation added.
The boundary of the spacetime $x^{\pm}$ and that of a Poincare $AdS_{3}$ $W^{\pm}$ is related by the conformal map
\be
W^{+}= e^{\f{2\pi}{\beta}( x^{+}-f(x^{+}))} \qquad W^{-}= e^{\f{2\pi}{\beta}( x^{-}-f(x^{-}))}
\ee
From the map one can compute the length of both extremal surfaces in the spacetime.
The contribution of the connected surface (\ref{eq:NBAS})  to holographic entanglement entropy   is 
\begin{align}
S_{c}(l_{1},l_{2},t)&= \f{c}{3}\log \left[\f{\beta}{2\pi a_{Q}}\sinh \f{\pi(l_{2}-l_{1})}{\beta} \right] \label{eq:iqec} \nonumber \\
&-\f{c\pi}{6\beta} \left[\left(f(l_{2}+t)-f(l_{1}+t) \right)+\left(f(l_{2}-t)-f(l_{1}-t) \right) \right] \nonumber \\ 
&+\f{c}{12} \left[f'(l_{1}+t)+f'(l_{2}+t)+f'(l_{1}-t)+f'(l_{2}-t) \right] \\ \nonumber  
\end{align}
 To derive the expression, we assume that inhomogeneity is small $ h(x) <<\beta$. 
The contribution  of the disconnected surface is 
\begin{align}
S_{dc}(l_{1},l_{2},t)&=\f{c}{3}\log \left[\f{\beta}{2\pi a_{Q}} \cosh \f{2\pi t}{\beta} \right] \label{eq:iqed} \nonumber \\
&-\f{c\pi}{6\beta}\left[\left(f(l_{1}+t)-f(l_{1}-t) \right)+\left(f(l_{2}+t)-f(l_{2}-t) \right) \right] \nonumber \\
&+\f{c}{12} \left[f'(l_{1}+t)+f'(l_{2}+t)+f'(l_{1}-t)+f'(l_{2}-t) \right] \nonumber \\   
\end{align}
Since we are assuming the fluctuation is infinitesimal $ h(x) <<\beta$, the critical time when the phase transition of both surface happen does not change.
By using (\ref{eq:backrelation}), we find an agreement between the CFT result (\ref{eq:inhee}) and the holograhic result.

One can rewrite the expression of entanglement entropy (\ref{eq:inhee}) into a suggestive form by making use of (\ref {eq:inhem}). When inhomogeneity is small, one can neglect $f'''(x \pm t)$ terms in (\ref {eq:inhem}). Then the variation of energy density is 
\be
\delta \langle  T_{tt}(x,t) \rangle =\f{c\pi}{6\beta^3}\left[ h(x+t)+ h(x-t) \right]
\ee
Since the last terms in (\ref{eq:iqec}), (\ref{eq:iqed}) can be neglected in high temperature limit $\beta \rightarrow 0$, The variation of 
entanglement entropy can be recast into the form,
\begin{align}
\delta S(l_{1},l_{2},t)&= \theta \left(\f{l_{2}-l_{1}}{2}-t\right) \beta  \int^{t}_{0} \left( \delta \langle T_{tt}(l_{1},s) \rangle + \delta \langle T_{tt}(l_{2},s) \rangle \right)ds \nonumber \\
&+\theta \left(t-\f{l_{2}-l_{1}}{2}\right)\beta \int^{l_{2}}_{l_{1}}  \delta \langle T_{tt}(s,t) \rangle ds
\end{align}
The second term of the expression comes from the first law of thermodynamics, since at late time entanglement entropy is reduced to thermal entropy in high temperature limit. The first term is specific to the quench. Holographically, this term comes from the fluctuation of the inside of the horizon.

\subsection{A finite inhomogeneous quench}
In the previous subsection, we consider  quenches  where inhomogeneity of energy density are infinitesimal. 
In this section we consider a quench where inhomogeneity  is finite.
We find for example following conformal map is interesting. 
\be
W^{\pm}=\left(\f{-1+\s{1+4e^{\f{2x^{\pm}}{\lambda}} }}{2}\right)^{\f{\pi \lambda}{\beta}} \label{eq:cmiq}
\ee

In the region $\f{x^{\pm} }{\lambda} \gg1$, one can approximate the map as $W^{\pm} \sim e^{\f{\pi x^{\pm}}{\beta}}$, thus the
corresponding bulk geometry is a black string metric with temperature $T=\f{1}{2\beta}$. On the other hand,   the region  
 $\f{x^{\pm} }{\lambda} \ll-1$ corresponds  to a black string geometry with $T=\f{1}{\beta}$. 

\begin{figure}
\begin{center}
\includegraphics[width=50mm]{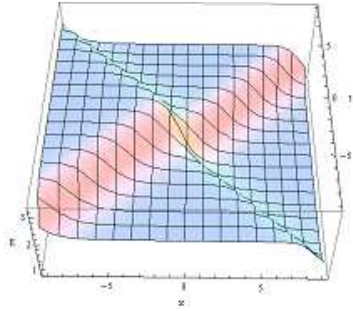}
\end{center}
\caption{The time volution of energy density $\langle T_{tt}(t,x) \rangle$ in the finite inhomogeneous quench.}
\label{fig:ef}
\end{figure}

In figure \ref{fig:ef} we plot the time evolution of the  energy density $\langle T_{tt}(t,x) \rangle$ in the quench.
In the figure one can see the third plateau in $|x|<|t|$. Below we will confirm that  
the region is described by a black string with $T=\f{3}{4 \beta}$ which is the 
average of temperature of two black strings on both sides.  
Thus in the bulk a black string with temperature $T_{1}= \f{1}{\beta}$ (below we  call it black string 1) and  a black string with temperature $T_{3}=\f{1}{2\beta}$(black string 3) are colliding at $x=0$ and $t=0$ , and a new black string with temperature $T_{2}=\f{3}{4\beta}$ is emerging (black string 2). The emergent black string expands in the speed of light form the origin. The final state of the process is a thermal equilibrium in terms of the emergent black string 2. The parameter
$\lambda$ is interpreted as the size of the intermediate regions between each of two black strings.

One can compute the contribution of both connected and disconnected surface $S_{c}(l_{1},l_{2},t)$, $S_{dc}(l_{1},l_{2},t) $ to the holographic entanglement entropy by using  general formulas 
(\ref{eq:ceegen}), (\ref{eq:dceegen}) in the Appendix and the conformal map (\ref{eq:cmiq}). Since exact expressions are complicated and do not illuminating, we do not depict them. The late time limit of the $S_{c}(l_{1},l_{2},t)$ is,
\begin{align}
S_{c}(l_{1},l_{2},t)&\rightarrow \f{c}{6} \log \left[ (\f{\beta}{2\pi a})^{2} \sinh \f{\pi}{\beta}(l_{2}-l_{1})\sinh \f{\pi}{2\beta}(l_{2}-l_{1}) \right] \quad t \rightarrow \infty \nonumber \\ 
&\sim \f{c\pi}{4\beta}(l_{2}-l_{1}),
\end{align}
which is equal to thermal entropy with temperature $T=\f{3}{4\beta}$.
On the other hand, the final state of the process is  the emergent black string (black string 2). Thus as we have advertised, the temperature of the emergent black string is determined to be  $T_{2}=\f{3}{4\beta}$.

In figure (\ref{fig:iqee}) we plot $S_{c}(l_{1},l_{2},t)$ and $S_{dc}(l_{1},l_{2},t)$ with $l_{1} <0 < l_{2}$.  Let us first discuss  the time dependence of 
 $S_{c}(l_{1},l_{2},t)$ of the connected surface.  When $t< \min \{|l_{1}|,l_{2}\}$, $S_{c}(l_{1},l_{2},t)$ remains constant.
Then in $\min \{|l_{1}|,l_{2}\} \leq t< \max \{|l_{1}|,l_{2} \}$ it increases linearly in time.
 When $ t \geq \max \{|l_{1}|,l_{2} \}$, it again becomes constant. 
This behavior can be interpreted in terms of the following bulk picture.
In the high temperature limit $\f{\lambda}{\beta} \ll 1$, each intermediate regions which connects each black strings can be ignored. In the limit, if we take a time slice $t=t_{0}$, each  black string  is extending between $-\infty <x<t_{0}$ (string 1),  $-t_{0} <x <t_{0}$ (string 2), $t_{0} < x < \infty$ respectively. Apart from junction points black strings are in local equilibrium with their own temperature.

When $t< \min \{|l_{1}|,l_{2}\}$, two end points of the emergent black string (string2) are located inside of the interval $[l_{1},l_{2}] $, see figure 4. The $S_{c}(l_{1},l_{2},t)$ of a BTZ black string with inverse temperature $\beta$ is given by  
\be
S_{c}(l_{1},l_{2},t)_{BTZ}= \f{c\pi}{3\beta} (l_{2}-l_{1}),
\ee
in the high temperature limit $\beta \ll  (l_{2}-l_{1})$. The  $S_{c}(l_{1},l_{2},t)$ in $t< \min \{|l_{1}|,l_{2}\}$ is thus obtained by summing  contributions form three black strings.
\begin{align}
 S_{c}(l_{1},l_{2},t) &=\f{c\pi}{3} \left[\f{|l_{1}|-t}{\beta} +\f{3}{4\beta}2t+\f{l_{2}-t}{2\beta} \right] \nonumber \\
&=\f{c\pi}{3} \left( \f{|l_{1}|+l_{2}}{2\beta}\right). \label{eq:iqce}
\end{align}
This explains the time independence of $ S_{c}(l_{1},l_{2},t)$ in $t< \min \{|l_{1}|,l_{2}\}$. From the plot one can also confirm that the value of the constant is actually given by (\ref{eq:iqce}).

When $\min \{|l_{1}|,l_{2}\}  \leq t< \max \{|l_{1}|,l_{2} \}$, one of the end point of the 
emergent black string is located at outside of the interval $[l_{1},l_{2}]$ due to the expansion.
Therefore $S_{c}(l_{1},l_{2},t)$ is 
\be
S_{c}(l_{1},l_{2},t)=\f{c\pi}{3} \left[\f{t}{4\beta}+\f{1}{4\beta}(2l_{2}+3|l_{1}|) \right].  \label{eq:iqcem}
\ee
 Therefore it grows linearly in time with coefficient $\f{c\pi}{12\beta}$. We check can this in figure \ref{fig:iqee}.

When $t \geq  \max \{|l_{1}|,l_{2} \}$, both of the end points of string 2 are located out side of the interval $[l_{1},l_{2}] $, thus $S_{c}(l_{1},l_{2},t)$ is thermalized
by the black string.
\be
S_{c}(l_{1},l_{2},t)= \f{c\pi}{4\beta}(l_{2}-l_{1}).
\ee

Next we consider  the behavior of $S_{dc}(l_{1},l_{2},t)$  of the disconnected surface.
Although it increases linearly in time, its coefficient $\f{\p S_{dc}}{\p t}$ jumps
at $t=\min \{|l_{1}|,l_{2}\} $ and $t=\max \{|l_{1}|,l_{2}\}$, see Figure \ref{fig:iqee}. 
The behavior can be understood again in terms of the expansion of the emergent black string .

We know that in a BTZ black string the disconnected surface is consist of two part, and each part is located at $x=l_{1}$or$x= l_{2}$. Each of them contribute to holographic entanglement entropy  as
  \be
S_{dc}(l_{1},l_{2},t)_{BTZ} =\f{\pi c}{3\beta}t.
\ee

When  $t< \min \{|l_{2}|,l_{1}\}$, one part of the disconnected surface probe the
 string 1 and the other probe the string 3 respectively. Thus the total contribution is 
\be
S_{dc}(l_{1},l_{2},t)=\f{\pi c}{2\beta}t.\label{eq:dcinh}
\ee
When  $\min \{|l_{1}|,l_{2}\} \leq t< \max \{|l_{1}|,l_{2}\}$, we have two cases. When $|l_{1}|<l_{2}$, since one  part of the
surface probe the string 2 and another probe the string 3, the total contribution is
\be
S_{dc}(l_{1},l_{2},t)= \f{\pi c}{12\beta}\left(5t +|l_{1}| \right).\label{eq:iqdcm} 
\ee
In the expression we add a constant term for continuity of $S_{dc}$ at $t=\min \{|l_{2}|,l_{1}\}$.
When $|l_{1}|\geq l_{2}$ they probe the string 2 and the string 1, thus
\be
S_{dc}(l_{1},l_{2},t)= \f{\pi c}{12\beta}\left(7t -l_{2} \right). \label{eq:iqdcmm} 
\ee
comparing them with (\ref{eq:dcinh}) we conclude  there is a jump in $\f{\p S_{c}}{\p t}$
at $t=\min \{|l_{1}|,l_{2}\}$

after $t=\max \{|l_{1}|,l_{2}\}$ both part probe the black string 2, then
\begin{eqnarray}
S_{dc}(l_{1},l_{2},t)=\left\{ \begin{array}{ll}
 \f{\pi c}{12\beta}\left(6t+5l_{2}+|l_{1}| \right)& (|l_{1}|<l_{2}) \\
 \f{\pi c}{12\beta}\left(6t+7|l_{1}| -l_{2}\right)& (l_{2}<|l_{1}|) \\
\end{array} \right.
\end{eqnarray}
again we find a jump at $t=\max \{|l_{1}|,l_{2}\}$.

Next we consider the phase transition between two surfaces. When $|l_{1}|<l_{2}$, from Figure \ref{fig:iqeee}  we see this happens
when (\ref{eq:iqcem}) and (\ref{eq:iqdcm}) become equal. Thus we find the critical time is
\be
t_{c}=\frac{l_{2}-l_{1}}{2} ,
\ee
which is the half of the size of the subsystem. We can check that this is also true when $l_{2}<|l_{1}|$. The expression agrees with that of derived from BTZ black string dual to the global  quench. In the CFT quasiparticle picture, the result is anticipated.
We also compare the gravity result for the quench and CFT result in figure 6, finding a nice agreement.

\clearpage
\begin{figure}
\begin{minipage}{0.3\hsize}
\begin{center}
   \includegraphics[width=40mm]{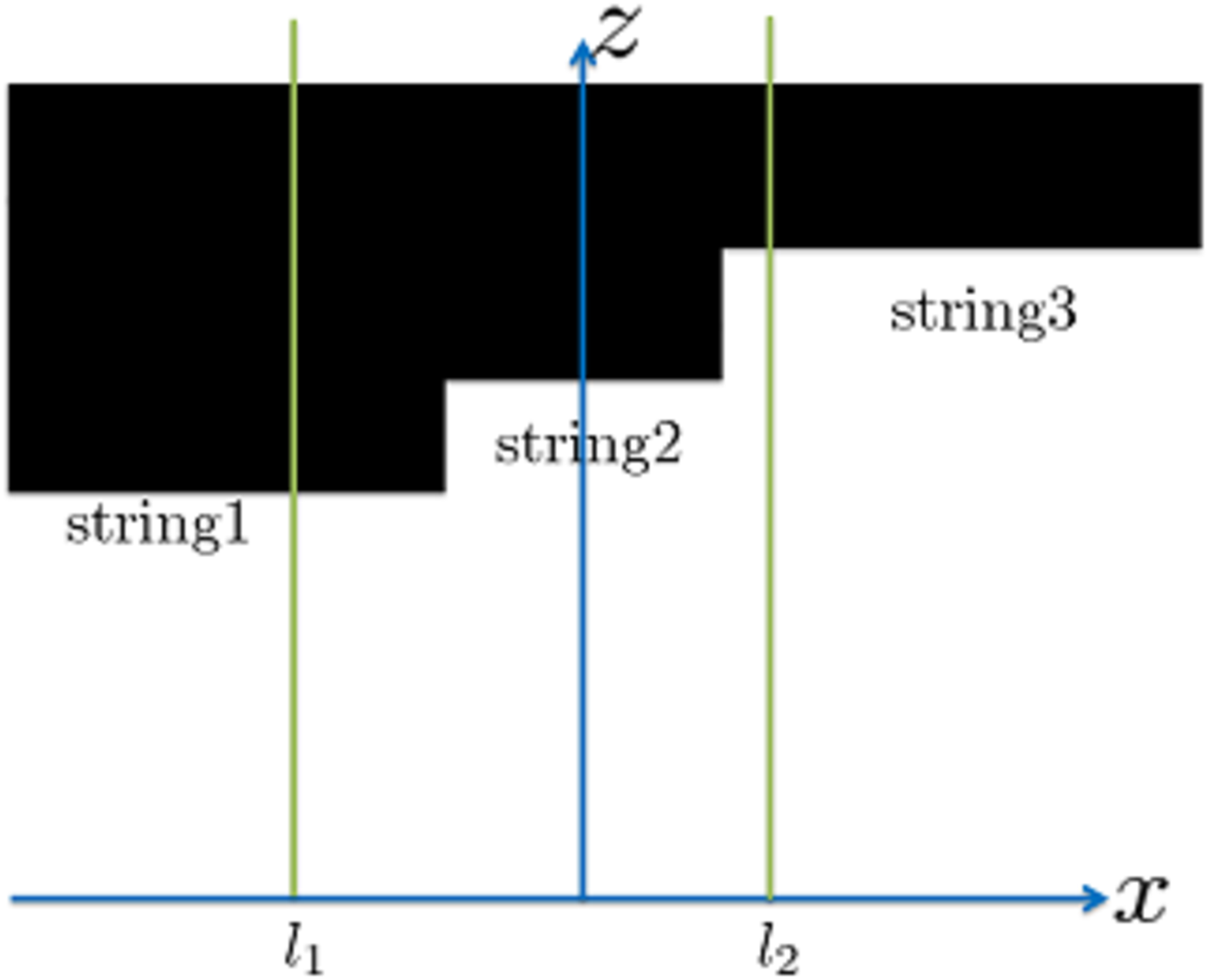}
\end{center}
\end{minipage}
\begin{minipage}{0.3\hsize}
\begin{center}
\includegraphics[width=40mm]{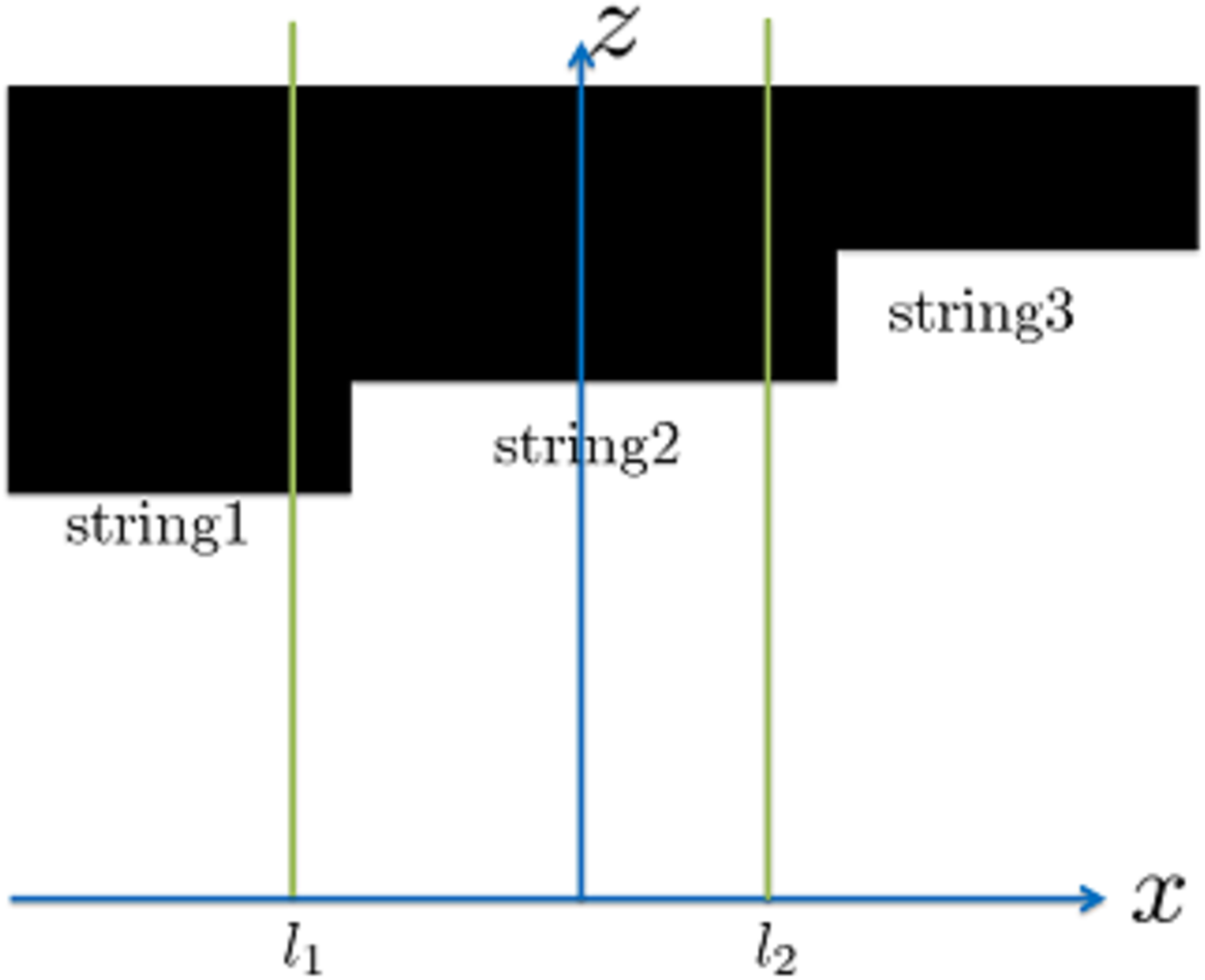}
\end{center}
\end{minipage}
\begin{minipage}{0.3\hsize}
\begin{center}
   \includegraphics[width=40mm]{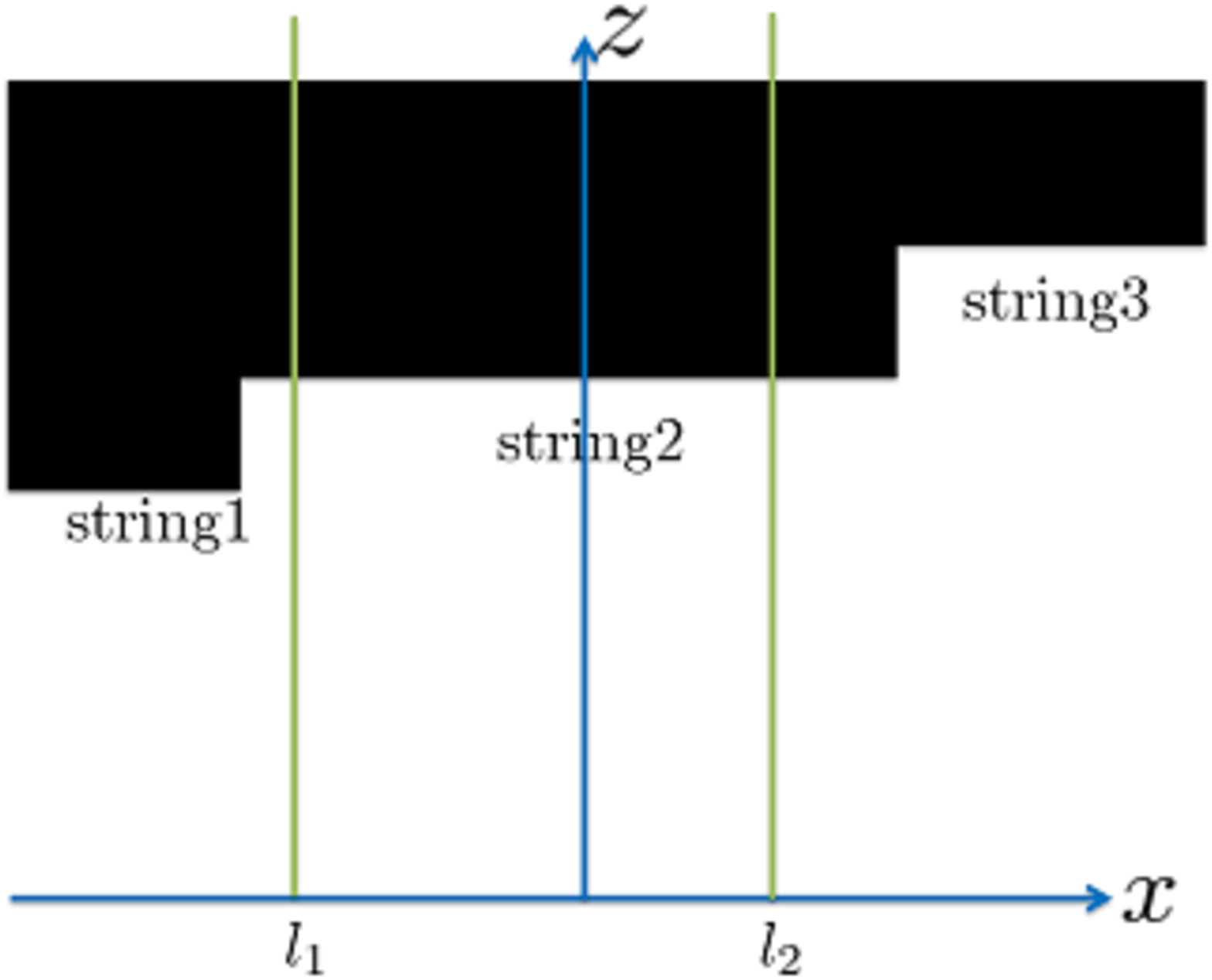}
\end{center}
\end{minipage}
\caption{Sketch of the fusion of two black strings (string1 and string3) into string 2.
Left: When $t<\min \{|l_{1}|,l_{2}\}$, the string 2 is contained in the bulk extension of the subsystem $[l_{1},l_{2}]$. Middle:When
$\min \{|l_{1}|,l_{2} \} \leq t<\max \{|l_{1}|,l_{2}\}$, one end point of the string 2  is located outside of the subsystem and the other is in the bulk extension of the subsystem.
Right: When $t \geq \max \{|l_{1}|,l_{2}\}$ both end points are located out side the bulk extension of the subsystem.}
\label{fig:surface}
\end{figure}
\begin{figure}
\begin{minipage}{0.5\hsize}
\begin{center}
   \includegraphics[width=50mm]{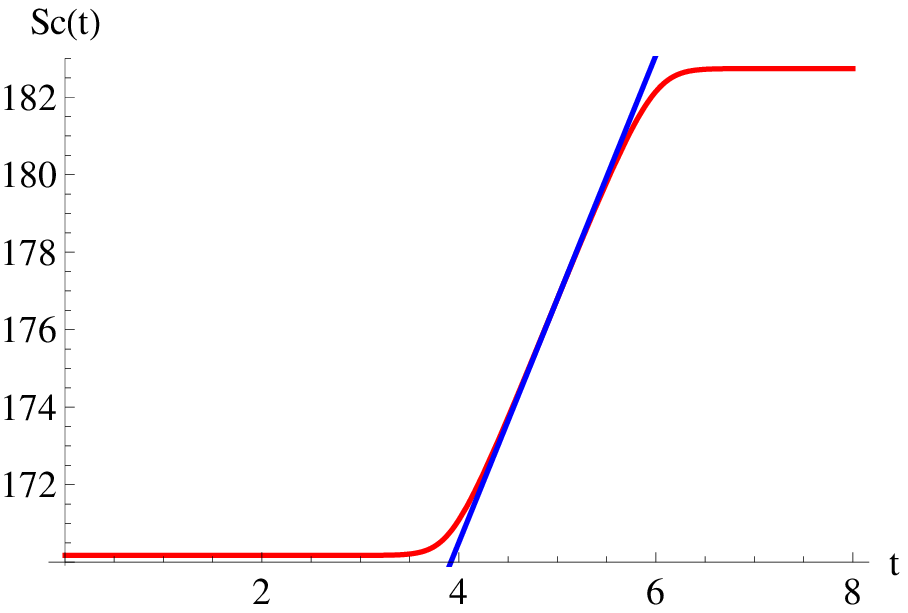}
\end{center}
\end{minipage}
\begin{minipage}{0.5\hsize}
\begin{center}
\includegraphics[width=50mm]{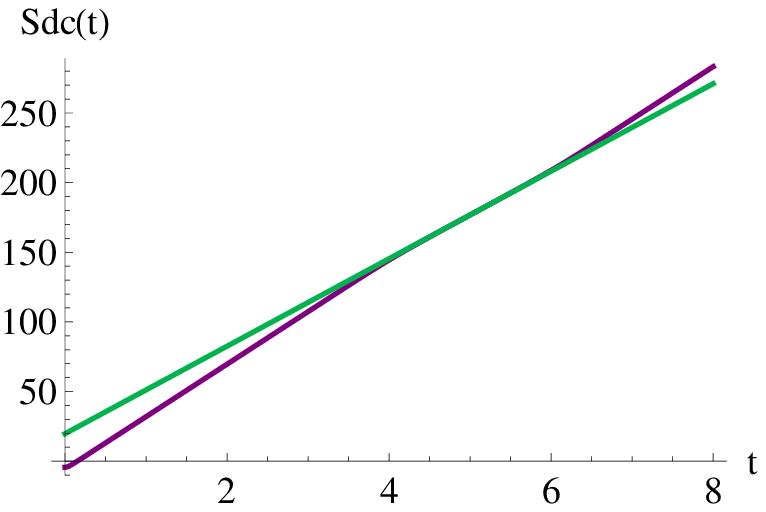}
\end{center}
\end{minipage}
\caption{Left: Plot of the length of the connected  surface as a function of time in the finite inhomogeneous quench (red). We also plot (\ref{eq:iqcem}) (blue).  
Right:Plot of the length of the connected  surface as a function of time in the finite inhomogeneous quench(green). We also plot (\ref{eq:iqdcm}) 
We take  $l_{1}=-4,l_{2}=6
,\lambda=\f{1}{4},\beta=\f{1}{4}$.}
\label{fig:iqee}
\end{figure}

\begin{figure}
\begin{minipage}{0.5\hsize}
\begin{center}
   \includegraphics[width=50mm]{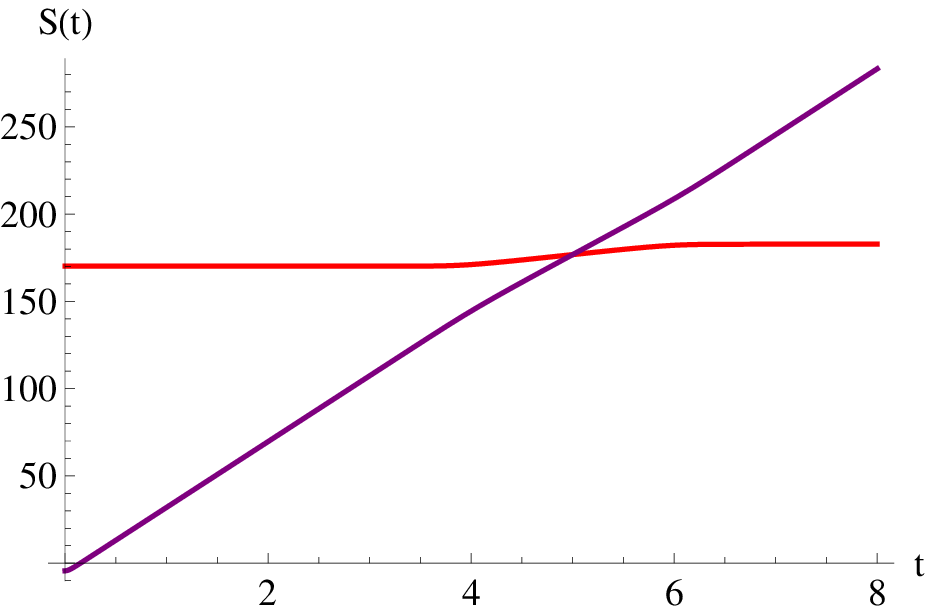}
\end{center}
\end{minipage}
\begin{minipage}{0.5\hsize}
\begin{center}
\includegraphics[width=50mm]{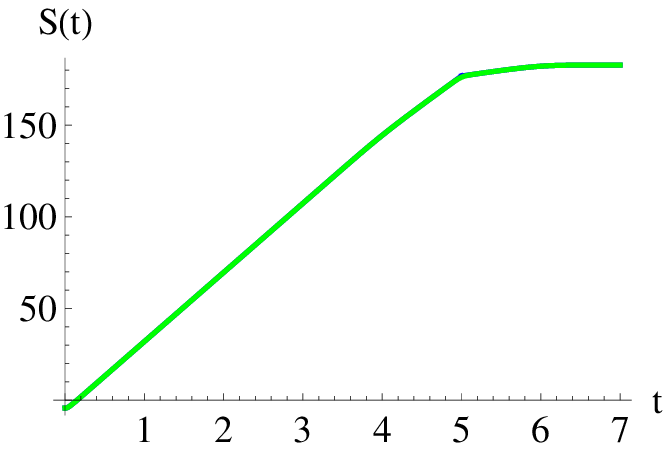}
\end{center}
\end{minipage}
\caption{Left: We compare the contribution of the disconnected surface (purple) and connected surface (red) to holographic entanglement entropy.  
Right: We compare the CFT result (green) and holographic result (blue), finding a nice agreement.
We take  $l_{1}=-4,l_{2}=6
,\lambda=\f{1}{4},\beta=\f{1}{4}$.}
\label{fig:iqeee}
\end{figure}

\clearpage

 \subsection{holographic dual of local quench}

In local quenches we heat up one point of the initial time slice. By these quenches 
an entangled pair of quasi particles are emitted from the point we heated up.
We can see the behavior from the one point functions of the stress energy tensor.
\be
\langle T_{++}\rangle=\f{c\epsilon^2}{8(x_{+}^2+\epsilon^2)^2} \qquad \langle T_{--} \rangle=\f{c\epsilon^2}{8(x_{-}^2+\epsilon^2)^2} .
\ee
The component of the stress tensor $\langle T_{++}\rangle$ have a peak at $x_{+}=0$ which is the world line
of the emitted leftmoving quasiparticle  In the high energy  limit $\epsilon \rightarrow 0$ the stress tensor localizes on the world line. One can also extract the world line of 
the rightmoving quasiparticle from $\langle T_{--} \rangle$.

Now we consider the time evolution of holographic entanglement entropy under the local quench. An attempt has been made in \cite{Nozaki:2013wia}. We take subsystem A to be $[l_{1},l_{2}]$ and assume $l_{2}>0$. 

The contribution from the disconnected surface to holographic entanglement entropy is derived by substituting the analytic continuation of the conformal map (\ref{eq:maplq}) into the general formula  (\ref{eq:ceegen}) in the Appendix. The result is 
\be
S_{dc}= \log \f{\left(2l_{1} +\s{(l_{1}+t)^2+\epsilon^2} +\s{(l_{1}-t)^2+\epsilon^2}\right)\left(2l_{2} +\s{(l_{2}+t)^2+\epsilon^2} +\s{(l_{2}-t)^2+\epsilon^2}\right)}{a_{LQ}^2 \s{\left(1+\f{l_{1}+t}{\s{(l_{1}+t)^2+\epsilon^2}} \right) \left(1+\f{l_{2}+t}{\s{(l_{2}+t)^2+\epsilon^2}} \right) \left(1+\f{l_{1}-t}{\s{(l_{1}-t)^2+\epsilon^2}} \right)\left(1+\f{l_{2}-t}{\s{(l_{2}-t)^2+\epsilon^2}} \right)}}.
\ee
In figure \ref{fig:lqee1} we plot the behavior of  the $S_{dc}(t)$.When $t<\min \{|l_{1}|,l_{2}\}$,  $S_{dc}(t)$ remains constant with the value 
\be
S_{dc}(0)=\frac{c}{6} \log \f{4|l_{1}|l_{2}}{a_{LQ}^2}, \label{eq:holde}
\ee
in the high energy limit $ \epsilon \rightarrow 0$. We can also see at $ t=l_{1}$ and $t=l_{2}$, the derivative $S'_{dc}(t)$ suddenly jumps. At late time $t>>1$, the $S_{dc}(t)$ keep increasing.
\be
S_{dc} \rightarrow \f{c}{3}\log \f{t^2}{\epsilon a_{LQ}} \quad t\rightarrow \infty . \label{eq:holdl}
\ee

The contribution of the connected surface is
\be
S_{c}(t)=\f{c}{6} \log \f{\left(l_{2}-l_{1} +\s{(l_{2}+t)^2+\epsilon^2} -\s{(l_{1}+t)^2+\epsilon^2}\right)\left(l_{2}-l_{1} +\s{(l_{2}-t)^2+\epsilon^2} -\s{(l_{1}-t)^2+\epsilon^2}\right)}{a_{LQ}^2 \s{\left(1+\f{l_{1}+t}{\s{(l_{1}+t)^2+\epsilon^2}} \right) \left(1+\f{l_{2}+t}{\s{(l_{2}+t)^2+\epsilon^2}} \right) \left(1+\f{l_{1}-t}{\s{(l_{1}-t)^2+\epsilon^2}} \right)\left(1+\f{l_{2}-t}{\s{(l_{2}-t)^2+\epsilon^2}} \right)}}.
\ee

We also plot the time dependence of $S_{c}(t)$ in figure \ref{fig:lqee1}. At late time $t \rightarrow \infty$, it reduces to the usual Poincare $AdS_{3}$ value,
\be
S_{c} \rightarrow \f{c}{3} \log \f{|l_{1}-l_{2}|}{a_{LQ}}  \quad t \rightarrow \infty\label{eq:holcl}.
\ee
The early time behavior of $S_{c}(l_{1},l_{2},t)$ depends on the sign of $l_{1}$.  When $l_{1}>0$ it reduces to the Poincare  $AdS_{3}$ value,
\be
S_{c}(0)=\f{c}{3} \log \f{|l_{1}-l_{2}|}{a_{LQ}}. \label{eq:holcpe}
\ee

When $l_{1}<0$, it becomes
\be
S_{c}(0)=\f{c}{3} \log \f{2|l_{1}|l_{2}}{a_{LQ} \epsilon}. \label{eq:holcne}
\ee

\begin{figure}
\begin{minipage}{0.3\hsize}
\begin{center}
   \includegraphics[width=40mm]{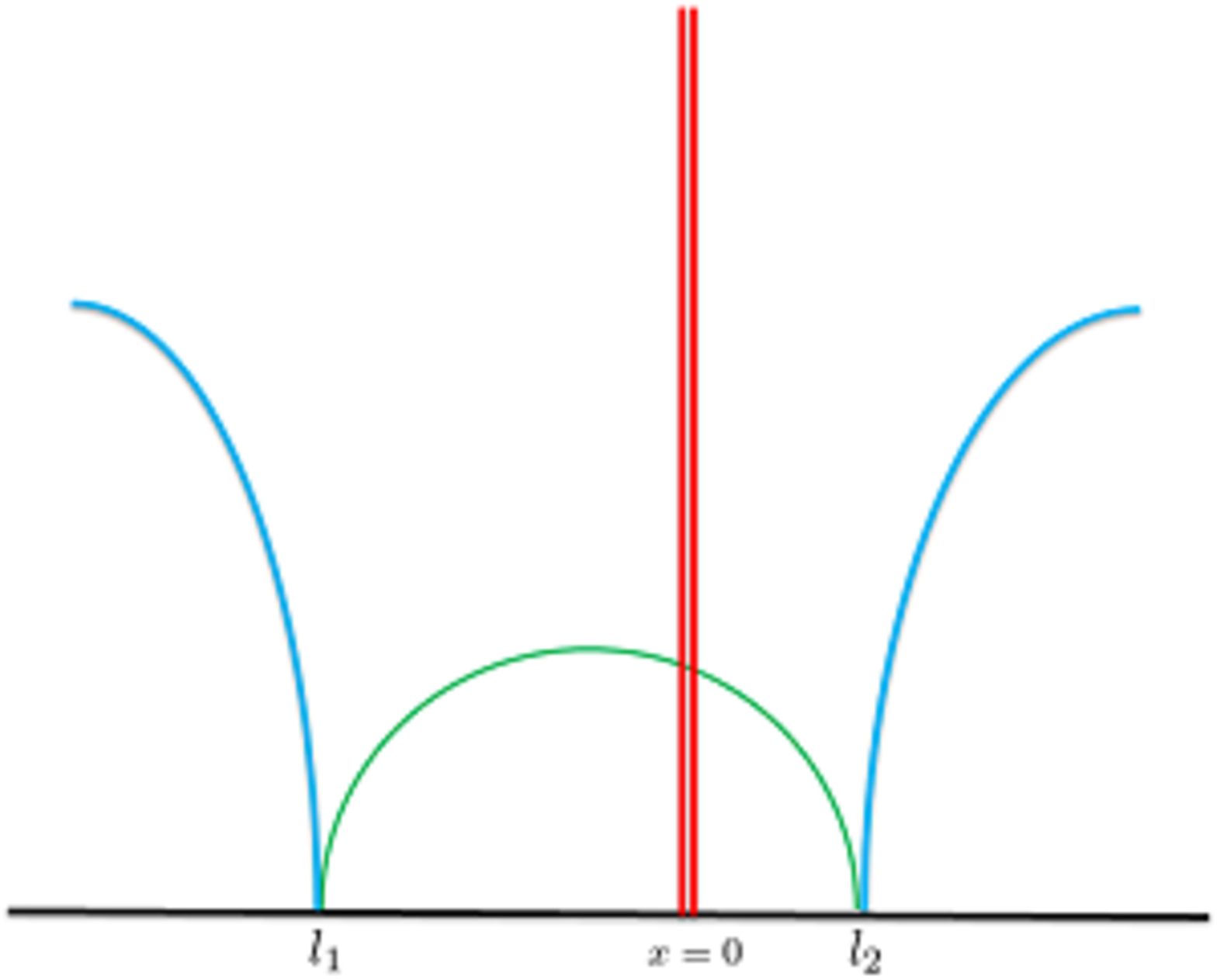}
\end{center}
\end{minipage}
\begin{minipage}{0.3\hsize}
\begin{center}
\includegraphics[width=40mm]{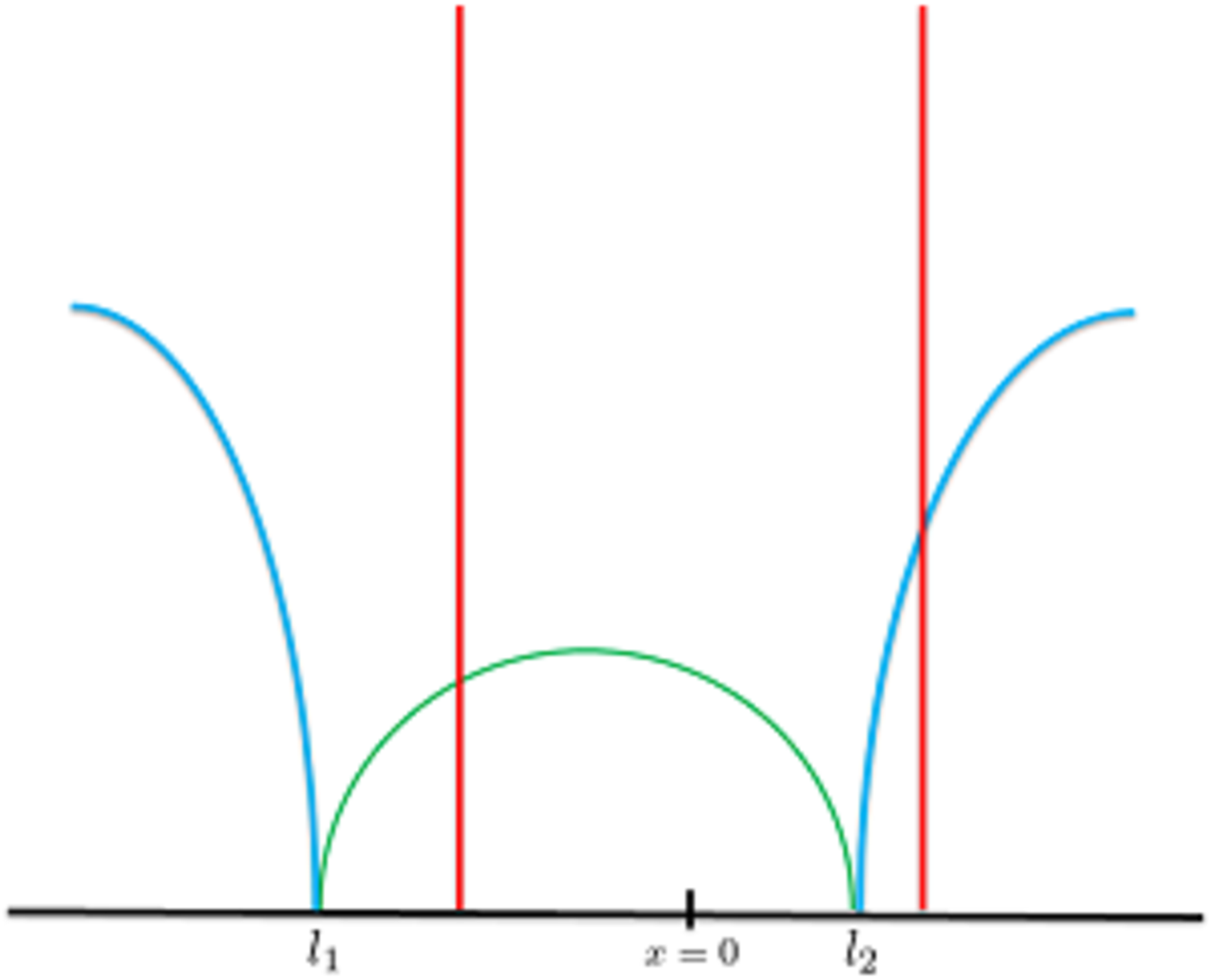}
\end{center}
\end{minipage}
\begin{minipage}{0.3\hsize}
\begin{center}
   \includegraphics[width=40mm]{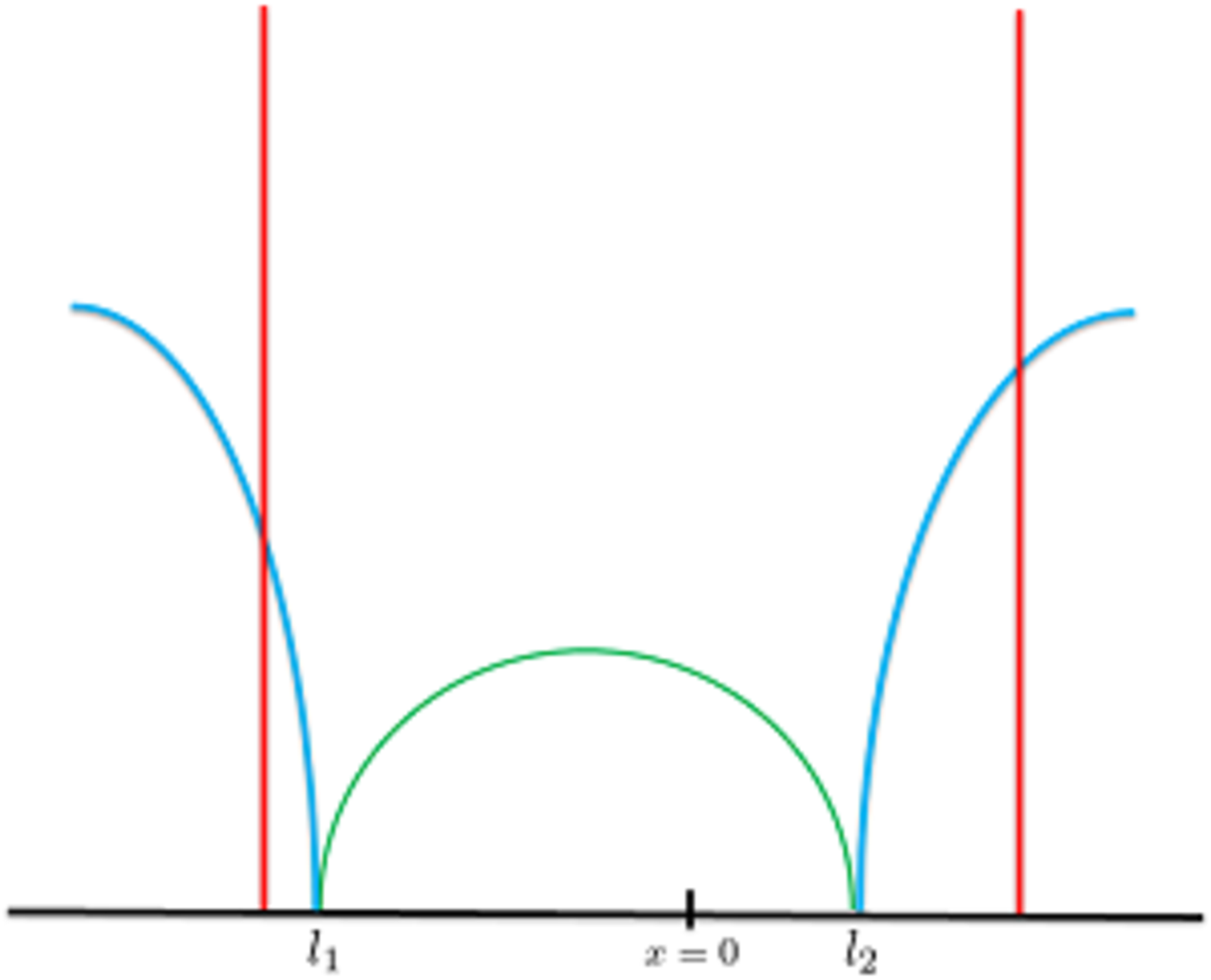}
\end{center}
\end{minipage}
\label{fig:surface}
\caption{Sketch of the extremal surfaces in $AdS_{3}$ (green:connected,blue:disconnected) and pulse (red line).
Left: When $t < \min \{|l_{1}|,l_{2}\}$, both  pulse intersect with the connected surface. Middle:When
$\min \{|l_{1}|,l_{2} \} \leq t<\max \{|l_{1}|,l_{2}\}$ one of the pulse intersect with the connected surface and the other intersect with the disconnected surface.
Right: $t \geq \max \{|l_{1}|,l_{2}\}$ both pulse intersect with the disconnected surface.}
\label{fig:catoo}
\end{figure}

\begin{figure}
\begin{minipage}{0.5\hsize}
\begin{center}
   \includegraphics[width=50mm]{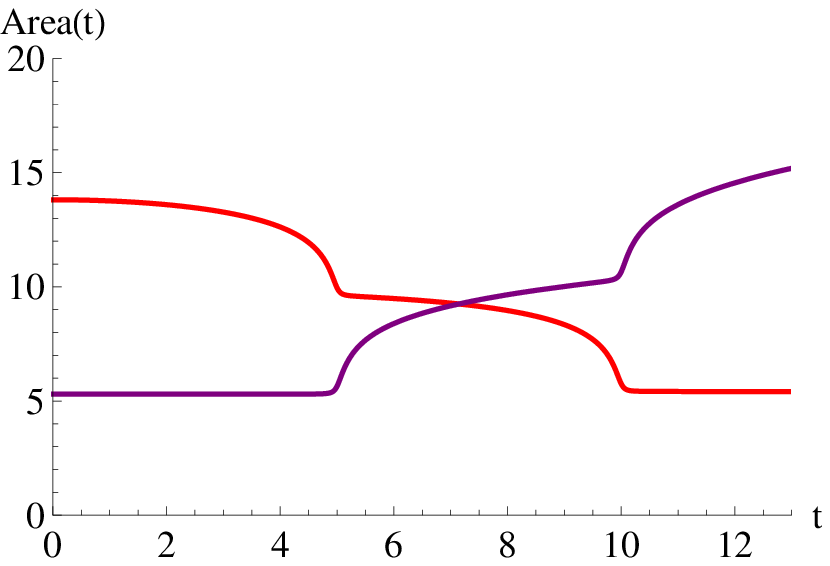}
\end{center}
\end{minipage}
\begin{minipage}{0.5\hsize}
\begin{center}
\includegraphics[width=50mm]{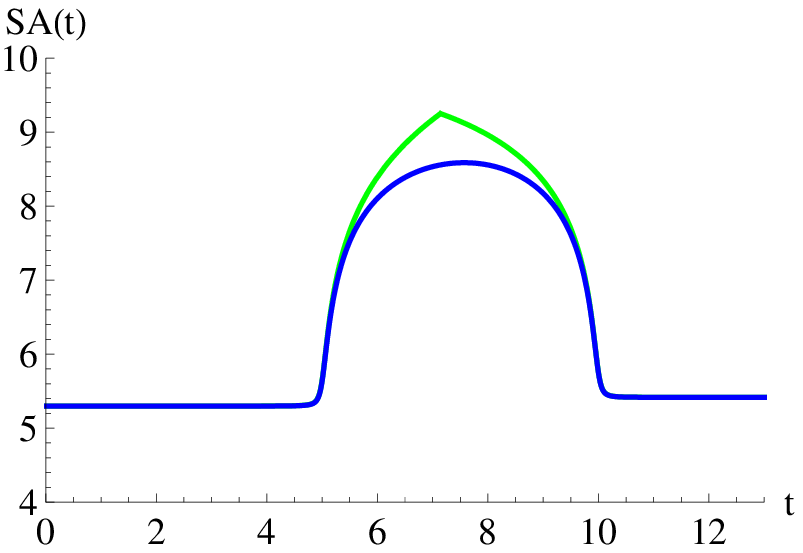}
\end{center}
\end{minipage}
\caption{Left:Plot of the length of the both surface as a function of time(red:connected, purple:disconnected).  We take $l_{1}=-5,l_{2}=10
,\epsilon=\f{1}{10}$. 
Right: comparison with CFT result(green: bulk, blue:CFT). }
\label{fig:lqee1}
\end{figure}

There is an intuitive interpretation of  the time 
dependence of $S_{c}(t)$ and $S_{dc}(t)$. In the high energy limit $\epsilon \rightarrow 0$, the expectation value of the stress tensor  localize near the pulse $x_{+}=0$, $x_{-}=0$
  ie, $\langle T_{++} \rangle \propto \delta ( x_{+})$ and $\langle T_{--} \rangle \propto \delta ( x_{-})$.  From (\ref{eq:met}), we see that the metric of the bulk is modified from that of  the Poincare $AdS_{3}$ only near the 
pulse. Note that the Poincare $AdS_{3}$ is different from the one \ref{eq:poincare} where the spacetime boundary is located at $X=0$. The pulse can be regarded as a bulk extension of the quasi particle pair. Extremal surfaces of the bulk can be regarded as  extremal surfaces of the Poincare $AdS_{3}$ with slight modification in the vicinity of the pulse.  As an intersection of an extremal surface and the pulse happen deeper and deeper in the bulk, the change of the length of the surface become larger and larger  because while the pulse significantly 
changes the IR metric of the bulk, but it doesn't change the UV metric which is fixed to be asymptotically $AdS_{3}$.  We notice the change of the length is positive from the explicit result. The figure \ref{fig:catoo}  shows the location of  two extremal surfaces in the Poincare $AdS_{3}$ together with the pulse at various time. Let us discuss the $l_{1}<0<l_{2}$ case as a specific example. When $0<t <\min \{|l_{1}|,l_{2} \}$, both of leftmoving and rightmoving  pulse intersect with the connected surface. Since  locations of  intersections gradually close to the boundary, the length of the connected surface is decreasing as time increases.
At  $\min \{|l_{1}|,l_{2} \} \leq t <\max \{|l_{1}|,l_{2} \}$, one part of the pulse intersects with the connected surface and the other with the disconnected surface. Since the intersection 
of the part of the pulse and the disconnected surface gradually leaves the boundary, the length  of the disconnected surface  is increasing.  At $t \geq \max \{|l_{1}|,l_{2} \}$, both part of the pulse intersect with the disconnected surface, thus the length of the disconnected surface is increasing but the length of the connected surface remains constant.
Sudden changes of the length of both surfaces at $t=l_{1}$ and $t= l_{2}$ in the figure  also support our intuitive picture. In $0<l_{1}<l_{2}$ case, one can also make a similar argument for the behavior of the both surfaces.

When $l_{1}\leq 0<l_{2}$, since the length of the disconnected surface is monotonically increasing in time, and the length of the connected surface is  monotonically decreasing,
there is a phase transition between two surfaces in general.  Because of inhomogeneity of the quench,
 The critical time $t_{c}$ when the phase transition happen depends on how we take a subsystem A. For example, when the subsystem  A is given by $[0,l]$, $t_{c}=l/4$
in high energy limit $\epsilon \rightarrow 0$. Note that the phase transition is crucial to the get correct early time behavior of  entanglement entropy holographically.
In figure \ref{fig:lqee1} we compare the CFT result and the holographic result for $l_{1}<0<l_{2}$ case. By comparing expressions (\ref{eq:holde}),(\ref{eq:holcl}) and (\ref{eq:lqcft1}),(\ref{eq:lqcft3})  we find both early time and late time behavior agree.
However near the critical time $t_{c}$, two results are different. The reason of the difference  come from the fact that in CFT side, we  approximate the two point function  on the upper half plane by (\ref{eq:2point}) to get a universal result. 

In figure \ref{fig:lqee1} we plot the time dependence of the both length of surfaces for $0<l_{1} <l_{2}$ case. In the case, the area of the connected surface is always smaller, and there is no
phase transition. By comparing it to the CFT result, we find that although the late time behavior agree (\ref{eq:holcl}) (\ref{eq:lqcft1}) , but early time does not (\ref{eq:holcpe}) (\ref{eq:lqcft2}). The reason of the difference is similar as the previous case. 

\clearpage

\begin{figure}
\begin{minipage}{0.5\hsize}
\begin{center}
   \includegraphics[width=50mm]{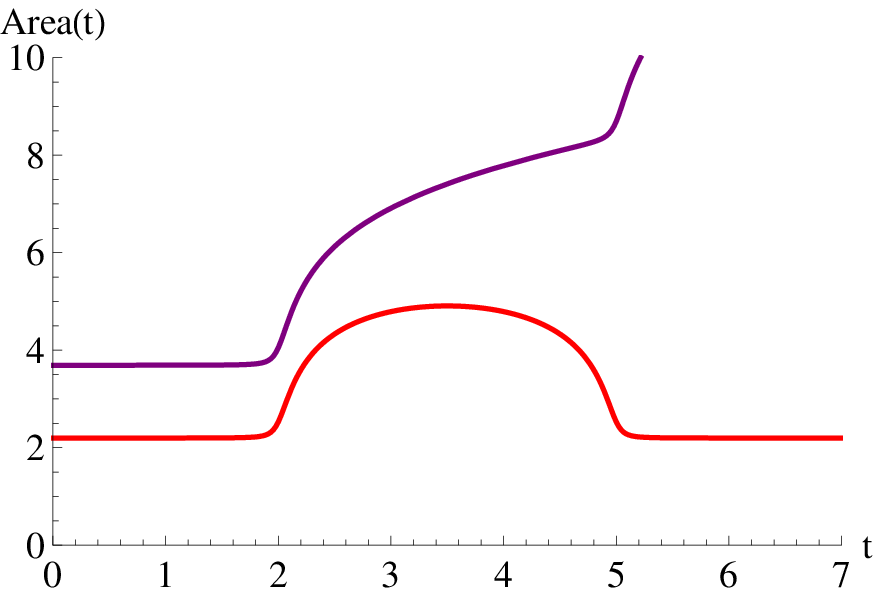}
\end{center}
\end{minipage}
\begin{minipage}{0.5\hsize}
\begin{center}
\includegraphics[width=50mm]{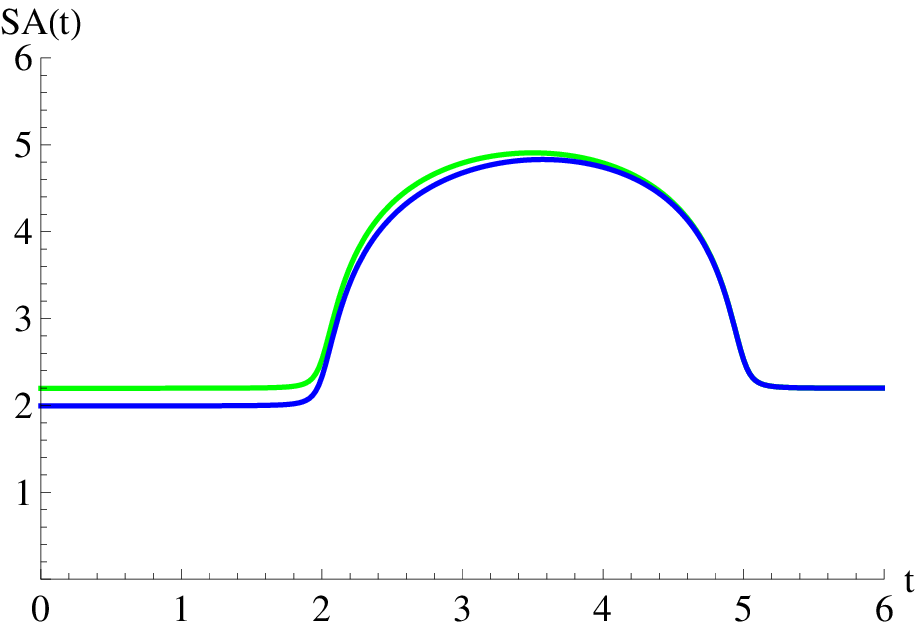}
\end{center}
\end{minipage}
\caption{Left:Plot of the length of the both surface as a function of time( red:connected, purple:disconnected).  We take $l_{1}=2,l_{2}=5,\epsilon=\f{1}{10}$. 
Right: comparison with CFT result(green: bulk, blue:CFT). }
\label{fig:lqee2}
\end{figure}
\clearpage

\section{Discussion}

In the paper we discussed a  holographically  realization of   quantum quenches of 2d CFTs.
In the construction we introduced a spacetime boundary into bulk geometries. The spacetiem boundary is the gravity counterpart of a boundary
state in CFTs\cite{Takayanagi:2011zk,Karch:2000ct,DeWolfe:2001pq}, which represents the initial condition of the quantum quench we consider. The location of the spacetime boundary in a bulk geometry is determined by pulling back the probe brane in a Poincare $AdS_{3}$ by the bulk extension of the associated Lorentzian conformal map.  

The introduction of a spacetime boundary means that when we compute holographic entanglement entropy,
we need to consider an extremal surface which ends on the spacetime boundary as well as boundary of the subsystem, 
in addition to conventional connected surface which only ends on the boundary of the subsystem. We called the surface as disconnected surface.
  We saw phase transitions between the connected surface and the disconnected surface explain  the whole time evolution of entanglement entropy in a quantum quench holographically.
Especially the length of the disconnected surface captures the early time behavior of entanglement entropy in most of all cases. 
We also found that when subsystem  is an infinite interval, the time evolution of entanglement entropy is entirely explained by the length of the disconnected surface in all quantum quenches.
 
We consider several examples as applications of the construction. We saw that a particular inhomogeneous quench  is dual to a spacetime in which  two black  strings merge  into the third black string. Except junction points,they are in local equilibrium and the temperature of these strings explain the behavior of entanglement entropy.   
We address the relation between the local quench in 2d CFTs and shock wave geometry. We saw that location of the shock wave determines the area of both surfaces.

\section*{Acknowledgements}
We would like to thank T.Takayanagi for discussions and reading the manuscript carefully.
T.U. is supported by World Premier International Research Center Initiative (WPI
Initiative), MEXT, Japan, and by JSPS Research Fellowships for Young Scientists.

\newpage

\section*{Appendix A }
here we summarize the general formulas of quantum quenches. As discussed in 
section 2, a quantum quench is specified by an associated Riemann surface $\Sigma$ with
boundary or the conformal map which maps $\Sigma$ into half plane. Let $W^{\pm}(x^{\pm})$ be the corresponding Lorentzian map derived by a Wick rotation.

The time evolution of entanglement entropy in CFT side is given by 
\begin{align}
S_{CFT}(l_{1},l_{2},t)&=\f{c}{6}\log \left[W^{+}(l_{1}+t)-W^{+}(l_{2}+t) \right]\left[W^{-}(l_{1}-t)-W^{-}(l_{2}-t) \right] \label{eq:cfteegen}\nonumber \\
&+\f{c}{6}\log \left[W^{+}(l_{1}+t)+W^{-}(l_{1}-t) \right]\left[W^{+}(l_{2}+t)+W^{-}(l_{2}-t) \right] \nonumber \\
&- \f{c}{6}\log \left[W^{+}(l_{1}+t)+W^{-}(l_{2}-t) \right]\left[W^{+}(l_{2}+t)+W^{-}(l_{1}-t) \right] \nonumber \\
&-\f{c}{6}\log a^2 \s{ \f{d W^{+}}{dx^{+}}(l_{1}+t) \f{d W^{-}}{dx^{-}}(l_{1}-t) \f{d W^{+}}{dx^{+}}(l_{2}+t) \f{d W^{-}}{dx^{-}}(l_{2}-t)}\nonumber  \\ ,
\end{align}
where $a$ denotes UV cut off of the CFT.
The contribution of the connected surface to the holographic entanglement entropy  is 
\begin{align}
S_{c}(l_{1},l_{2},t)&=\f{c}{6}\log \left[W^{+}(l_{2}+t)-W^{+}(l_{1}+t) \right]\left[W^{-}(l_{2}-t)-W^{-}(l_{1}-t) \right] \label{eq:ceegen}\nonumber \\
&-\f{c}{6}\log a^2 \s{ \f{d W^{+}}{dx^{+}}(l_{1}+t) \f{d W^{-}}{dx^{-}}(l_{1}-t) \f{d W^{+}}{dx^{+}}(l_{2}+t) \f{d W^{-}}{dx^{-}}(l_{2}-t)}\nonumber  \\.
\end{align}

The contribution of the disconnected surface is 
\begin{align}
S_{dc}(l_{1},l_{2},t)&=\f{c}{6}\log \left[W^{+}(l_{1}+t)+W^{-}(l_{1}-t) \right]\left[W^{+}(l_{2}+t)+W^{-}(l_{2}-t) \right] \label{eq:dceegen}\nonumber \\
&-\f{c}{6}\log a^2 \s{ \f{d W^{+}}{dx^{+}}(l_{1}+t) \f{d W^{-}}{dx^{-}}(l_{1}-t) \f{d W^{+}}{dx^{+}}(l_{2}+t) \f{d W^{-}}{dx^{-}}(l_{2}-t)}\nonumber  \\.
\end{align}

\end{document}